# Thermionic emission or tunneling? The universal transition electric field for ideal Schottky reverse leakage current in β-Ga$_2$O$_3$

Wenshen Li[1], Kazuki Nomoto[1], Debdeep Jena[1,2,3] and Huili Grace Xing[1,2,3,a)]
[1]School of Electrical and Computer Engineering, Cornell University, Ithaca, NY 14853, USA
[2]Department of Materials Science and Engineering, Cornell University, Ithaca, NY 14853, USA
[3]Kavli Institute at Cornell for Nanoscale Science, Cornell University, Ithaca, NY 14853, USA
(Dated: April 15, 2020)

[a)]Author to whom correspondence should be addressed: grace.xing@cornell.edu

The reverse leakage current through a Schottky barrier transitions from a thermionic-emission dominated regime to a barrier-tunneling dominated regime as the surface electric field increases. In this study, we evaluate such transition electric field ($E_T$) in β-Ga$_2$O$_3$ using a numerical reverse leakage model. $E_T$ is found to have very weak dependence on the doping concentration and barrier height, thus a near-universal temperature dependence suffices and is given by a simple empirical expression in Ga$_2$O$_3$. With the help of a field-plate design, we observed experimentally in Ga$_2$O$_3$ Schottky barrier diodes a near-ideal bulk reverse leakage characteristics, which matches well with our numerical model and confirms the presence of the transition region. Near the transition electric field, both thermionic emission and barrier tunneling should be considered. The study provides important guidance toward accurate design and modeling of ideal reverse leakage characteristics in β-Ga$_2$O$_3$ Schottky barrier diodes.

With a unique combination of an ultra-high breakdown electric field of ~8 MV/cm [1][2], a decent electron mobility of ~200 cm$^2$/V·s [3], an availability of melt-grown substrates [4] and controllable n-type doping [5], β-Ga$_2$O$_3$ is an attractive ultra-wide bandgap semiconductor material for applications demanding high power handling capability [6]. To date, promising performance in Ga$_2$O$_3$ power devices have been demonstrated, including Schottky barrier diodes with a breakdown voltage (BV) over 2 kV [7-9], and high-voltage power transistors [10-12].

Ga$_2$O$_3$ Schottky barrier diodes (SBDs) are highly versatile. They can not only be used as high-speed rectifiers for efficient power regulation [13-15], but also function as UV photodetectors [16][17]. In addition, a Schottky contact is a key device building block, offering gate control for metal-semiconductor field-effect transistors (MESFETs) [1], as well as serving potentially as a p-n junction replacement for high field management [18][19].

Central among all the aforementioned functionalities is the reverse blocking capability of the Schottky barrier, which is critically dependent on the reverse leakage current. In general, the ideal total reverse leakage current ($J_{R,tot}$) through a Schottky barrier consists of the transport of electrons both above and below the top of the barrier. The former mechanism is thermionic emission (TE), while the latter is barrier tunneling (BT), which comprises thermionic-field emission (TFE) and field emission (FE) [20][21]. It has been widely-recognized that, at a certain temperature, there should exist a transition voltage ($V_T$) or transition electric field ($E_T$), below which TE dominates, and above which BT dominates [20][21]. Knowledge about $V_T$ or $E_T$ is highly valuable, since it determines the appropriate bias or electric-field ranges for TE and BT models. However, due to the difficulty in calculating the tunneling current analytically near this transition region, there has not been a simple closed-from expression for $V_T$ or $E_T$. In previous studies, $V_T$ in β-Ga$_2$O$_3$ has been calculated numerically [22], but the dependence on the doping concentration, barrier height and temperature are very complicated. Also, the non-monotonic temperature dependence of $V_T$ is questionable. In this study, we first show from numerical calculation that, unlike $V_T$, $E_T$ is nearly independent of the doping concentration and the barrier height; furthermore, there exists a universal monotonic temperature dependence of $E_T$.

Experimentally, most of previous analysis on the reverse leakage current in Ga$_2$O$_3$ SBDs either uses TE model [17][23] or TFE model [24][25], without considering their respective appropriate ranges. We have previous observed and verified near-ideal bulk reverse leakage current in Ga$_2$O$_3$ SBDs fabricated on bulk substrates [18]. However, due to the high doping concentration, the transition electric field cannot be accessed. In this work, we fabricate SBDs on an n$^-$-Ga$_2$O$_3$ epitaxial layer, which allows us to access surface electric field around $E_T$. The edge leakage current is sufficiently suppressed with a field-plate structure.

For the calculation of the reverse leakage current, two effects on the shape of the Schottky barrier potential should be considered: image-force lowering (IFL) and doping effects, as illustrated schematically in Fig. 1(a) and (b), respectively. Including both effects in the calculation of the tunneling current is analytically intractable, thus we developed a numerical approach, as presented in our previous work [18]. Here, we recast the theoretical framework slightly differently, such that the thermionic emission current ($J_{TE}$) and barrier tunneling current ($J_{BT}$) are expressed separately.

The total reverse leakage current is a sum of $J_{TE}$ and $J_{BT}$:

$$J_{R,tot} = J_{BT} + J_{TE}. \qquad (1)$$

In the presence of both IFL and doping effects, the potential energy distribution of the Schottky barrier under a surface electric field $E$ is given by

$$\mathcal{E}_c(x) = e\phi_B - eEx - \frac{e^2}{16\pi\varepsilon_s x} + \frac{e^2 N_d x^2}{2\varepsilon_s}, \quad (2)$$

where $\phi_B$ is the barrier height, $N_d$ is the net donor concentration, and $\varepsilon_s = 10\varepsilon_0$ is the dielectric constant of β-Ga$_2$O$_3$ [26]. Here, the Fermi-level energy in metal ($\mathcal{E}_{Fm}$) is taken as the zero-energy level, as shown in Fig. 1(a). Due to the presence of IFL, the barrier lowered by $\Delta\phi = \sqrt{eE/(4\pi\varepsilon_s)}$, resulting in $\mathcal{E}_{c,max} = e(\phi_B - \Delta\phi)$. Assuming a transmission probability of unity for electrons with an energy $\mathcal{E}$ higher than $\mathcal{E}_{c,max}$, $J_{TE}$ is given by the familiar expression:

$$J_{TE} = A^* T^2 \exp\left(-\frac{e\phi_B - \Delta\phi}{k_B T}\right), \quad (3)$$

where $A^* = 4\pi m^* k_B^2 e/h^3$ is the Richardson constant. One the other hand, $J_{BT}$ can be expressed by [18][20]

$$J_{BT} = \frac{A^* T}{k_B} \int_{-\infty}^{\mathcal{E}_{c,max}} \mathcal{T}(\mathcal{E}) \cdot \ln\left[1 + \exp\left(-\frac{\mathcal{E} - \mathcal{E}_{Fm}}{k_B T}\right)\right] d\mathcal{E}, \quad (4)$$

where $\mathcal{E}$ is the electron energy, $\mathcal{T}(\mathcal{E})$ is the tunneling probability. Using a Wentzel–Kramers–Brillouin (WKB)-type approximation, $\mathcal{T}(\mathcal{E})$ can be written as [20]

$$\mathcal{T}(\mathcal{E}) = \left[1 + \exp\left(-\frac{2i}{\hbar} \int_{x_1}^{x_2} p(x) dx\right)\right]^{-1}, \quad (5)$$

where $p(x) = -i\sqrt{2m^*(\mathcal{E}_c(x) - \mathcal{E})}$, $x_1$ and $x_2$ are classical turning points where $\mathcal{E}_c(x) = \mathcal{E}$. A single effective mass $m^* = 0.31 m_0$ is adopted for both the Richardson constant and the tunneling effective mass due to the single-valley and near-isotropic nature of the conduction band [27][28].

Figure 1 shows the calculated $J_{R,tot}$ using the numerical model, as well as its constituent components ($J_{TE}$, $J_{BT}$). The transition electric field $E_T$ is defined at the surface electric field where $J_{TE} = J_{BT}$. Here, we have temporality neglected the doping effect to compare with the analytical TE and TFE models derived by Murphy and Good [20], which consider only the IFL. It can be seen that the calculated $J_{R,tot}$ from our numerical model agrees well with Murphy and Good's models within their respective applicable ranges, indicating that our numerical method is valid. It is worth noting that Murphy and Good's TE model actually includes both the TE and BT currents despite what the name suggests, thus it comes no surprise that a match with $J_{R,tot}$ is observed, rather than with $J_{TE}$.

With the numerical model established, the transition electric field $E_T$ can be calculated by equating $J_{TE}$ with $J_{BT}$. Figure 2(a) shows the calculated $E_T$ as a function of the net doping concentration ($N_D$-$N_A$) at different temperatures, under a barrier height of 1.2 eV. $E_T$ is primarily a function of temperature and has a very weak dependence on $N_D$-$N_A$. It is near constant when $N_D$-$N_A$<$10^{17}$ cm$^{-3}$, and only increases slightly (<0.08 MV/cm) when $N_D$-$N_A$ approaches $2\times10^{18}$ cm$^{-3}$, indicating that influence of the doping effect is near negligible. This illustrates the superiority of using surface electric field instead of the reverse bias as the variable to characterize the transition region, as $V_T$ would have large dependence on $N_D$-$N_A$ even with a constant $E_T$. The surface electric field at zero bias ($E_0$) is also calculated by the familiar expression: $E_0 = \sqrt{2(N_D - N_A)(eV_{bi,0} - k_B T)/\varepsilon_s}$, where $V_{bi,0}$ is the built-in potential at zero bias. If $E_0$ is larger than $E_T$, $J_{R,tot}$ would be dominated by barrier tunneling, as in the case of the SBD we reported in Ref. 18.

Knowing the weak dependence on the doping concentration, we calculate $E_T$ as a function of temperature without considering the doping effect, as shown in Fig. 2(b). Here, we examine the influence of the barrier height ranging from 0.5-2.0 eV. Interestingly, there is a negligible dependence on the barrier height from 100 K to 800 K. Below 100 K, there exists a sharp transition of $E_T$ to zero at some "transition" temperature, depending on the barrier height. The transition is unimportant since $E_T$ is already very low (<0.06 MV/cm). Also, with $N_D$-$N_A \geq 1\times10^{16}$ cm$^{-3}$ in practice, the value of $E_T$ around this region would already be lower than $E_0$, rendering it unobservable.

Due to the negligible dependence on the barrier height and very weak dependence on the doping concentration, we can describe $E_T$ in β-Ga$_2$O$_3$ with a near-universal empirical temperature dependence as obtained from a quadratic fitting to the numerical calculation in Fig. 2(b):

$$E_T = 0.70 \cdot T^2 + 780 \cdot T - 3.0 \times 10^4 \text{ V/cm}, \quad (6)$$

where $T$ is in the unit of K. Equation 6 is valid within the temperature range of 100-800 K and a barrier-height range of 0.5-2.5 eV, with a maximum error of 0.01 MV/cm for $N_D$-$N_A \leq 1\times10^{17}$ cm$^{-3}$, and 0.08 MV/cm for $N_D$-$N_A \leq 2\times10^{18}$ cm$^{-3}$.

To verify the existence of the transition region experimentally, we fabricated field-plated SBDs on a (001) Ga$_2$O$_3$ epitaxial wafer grown by halide vapor phase epitaxy (HVPE), as schematically shown in Fig. 3(a). The field plate length is designed to be 30 μm for the purpose of suppressing the electric field crowding at the anode edge. The cathode ohmic contact is based on Ti/Au (50/100 nm), while the anode Schottky contact is based on Ni/Au (40/150 nm). The fabrication process for the formation of cathode and anode contacts were the same as described in Ref. 29. After the anode formation, a 31-nm Al$_2$O$_3$ was deposited by atomic layer deposition (ALD) under 300 °C, acting as the dielectric for the field plate. Finally, a contact hole was etched, followed by the deposition of the field plate, which comprises a stack of Ni/Ti/Al/Pt (30/10/80/20 nm).

Temperature-dependent capacitance-voltage ($C$-$V$) measurements were performed on co-fabricated SBDs without the field plate. Figure 3(b) shows the extracted doping profile, which shows an average $N_D$-$N_A$ of ~$7\times10^{15}$ cm$^{-3}$ in the Si-doped n$^-$ drift layer. The $1/C^2$-$V$ plot is shown in the inset, from which $V_{bi,0}$ is extracted to be 1.10 V at 25 °C, and 0.92 V at 200 °C.

Figure 4(a) shows the measured temperature-dependent forward current-voltage ($I$-$V$) characteristics. Since the doping concentration is low, the depletion width at zero bias is ~0.4 μm. In this case, the TE model is inappropriate for the analysis of the forward $I$-$V$ characteristics, since the electron transport through the depletion region cannot be assumed ballistic, especially considering the mobility of Ga$_2$O$_3$ is not very high [3]. Therefore, we analyze the data with the thermionic emission-diffusion (TED) model [30][31], which considers the drift-diffusion transport in the depletion region. In the calculations, the temperature-dependent drift mobility model in Ref. 3 is used with the Hall factor considered. The constant of proportionality is adjusted to match the Hall mobility of 145 cm$^2$/V·s measured at 25 °C on a similar wafer [5].

The extracted barrier heights and ideality factors ($n$) from TED model at each temperature are plotted in Fig. 4(b). The ideality factor is 1.02 at 25 °C and decreases to below 1.01 beyond 100 °C, indicating a very good Schottky contact quality. The image-force controlled ideality factor limit ($n_{IF}$) is calculated to be 1.007 using the standard method [32]. It can be seen that the extracted ideality factor approaches $n_{IF}$ beyond 75 °C, further suggesting a near-ideal interface. Both the apparent barrier height as well as the barrier height after image-force (IF) correction (~0.026 eV) [32] was plotted. The IF-correct barrier height is around 1.20 eV.

To verify the effect of the field plate in suppressing the edge leakage current, temperature-dependent reverse $I$-$V$ measurements was performed on both non-field-plated and field-plated SBDs on the same wafer, as shown in Figs. 5(a) and (b) respectively. Without field plate, the SBDs exhibit a large, near temperature-independent reverse leakage current below 100 °C. As have been pointed out in our previous report [29], such a leakage behavior is characteristics of field-emission dominated edge leakage current due to electric-field crowding at the anode edge. On the other hand, the leakage current in SBDs with FP is much reduced, suggesting the FP structure is effective in suppressing the edge electric-field crowding. Note that between -200 V and -100 V, there still exists a very low level of leakage current at 25-75 °C that do not show much temperature dependence, suggesting that there is still some edge leakage current not completely eliminated. However, considering the very low magnitude (<10$^{-8}$ A/cm$^2$) and the weak temperature dependence, it will not significantly "pollute" the uniform bulk leakage current from 100 °C to 200 °C, which we will model using our numerical models.

Figure 6 shows the temperature-dependent reverse leakage current as a function of the surface electric field ($E$), i.e., $J$-$E$ characteristics. The reverse leakage characteristics from 100 °C to 200 °C can be well-fitted with the calculated *total* reverse leakage current ($J_{R,tot}$) using our numerical model (Eqs. 1-5), *with the barrier height as the only fitting parameter at each temperature.* The individual components of $J_{R,tot}$, i.e. $J_{TE}$ and $J_{BT}$, are also plotted in Fig. 6. While $J_{TE}$ matches with the measured data at the lower end of $E$ and $J_{BT}$ at the higher end of $E$, neither $J_{TE}$ nor $J_{BT}$ alone can capture well the field dependence throughout the entire electric-field range (0.07-0.69 MV/cm). These results strongly suggest the presence of the transition regime, where both $J_{TE}$ and $J_{BT}$ are important.

The validity of the fitting depends critically on the fact that the measured bulk reverse leakage current is near ideal. A good check would be comparing the barrier height values extracted from the reverse leakage characteristics with those extracted from other methods. Figure 7 shows such comparisons. The barrier height values extracted from forward $I$-$V$, reverse $I$-$V$ and $C$-$V$ measurements exhibit good agreements. These provide further evidence that the measured reverse leakage current is near ideal, which in turn corroborates the identifications of the transition region from the data fitting in Fig. 6. The barrier heights extracted from $C$-$V$ measurements is slightly larger than from the $I$-$V$ methods, especially at lower temperatures. This behavior is commonly observed (see, e.g., Ref. 17), and could be due to the presence of barrier height inhomogeneity [33] and/or uncertainty of the doping concentration close to the Schottky contact interface.

In conclusion, the transition electric field ($E_T$) separating the thermionic-emission and barrier-tunneling dominated reverse leakage regimes is calculated in β-Ga$_2$O$_3$ Schottky barrier diodes, by using a numerical reverse leakage model. $E_T$ is shown to have very weak dependence on the doping concentration and the barrier height. A near-universal empirical expression for $E_T$ is obtained, which is valid for wide temperature, doping, and barrier-height ranges in β-Ga$_2$O$_3$ SBDs. Experimentally, we confirmed the presence of the transition region in field-plated Ga$_2$O$_3$ SBDs, which show near-ideal bulk reverse leakage current well-matched with our numerical model. With the knowledge about the transition electric field, the long-standing confusion about whether thermionic emission model or tunneling model should be used is lifted: if the surface electric field is much lower than $E_T$, thermionic-emission model can be used; conversely, barrier-tunneling model should be employed. Near $E_T$, it is important to consider both models. These results are highly valuable for the design of functional Ga$_2$O$_3$ Schottky barriers that rely on the precise knowledge about the reverse leakage current.

DATA AVAILABILITY: The data that supports the findings of this study are available within the article.

ACKNOWLEDGEMENT: This work was supported in part by NSF DMREF 1534303 and AFOSR (FA9550-17-1-0048, FA9550-18-1-0529), and carried out at Cornell Nanoscale Facility and CCMR Shared Facilities sponsored by the NSF NNCI program (ECCS-1542081), MRSEC program (DMR-1719875) and MRI DMR-1338010.

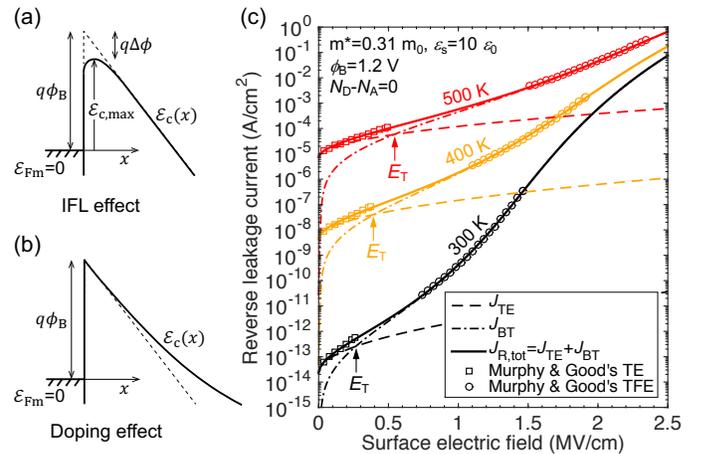

Fig. 1 Schematic illustrations of (a) the image-force lowering (IFL) effect, and (b) the doping effect. (c) Calculated total reverse leakage current ($J_{R,tot}$) as a function of the surface electric field in Ga$_2$O$_3$ Schottky barrier diodes (SBDs) using our numerical model, showing excellent agreements well with Murphy

and Good's analytical models [20]. The transition electric field ($E_T$) is illustrated at the cross-over point between the thermionic-emission current ($J_{TE}$) and the barrier-tunneling current ($J_{BT}$).

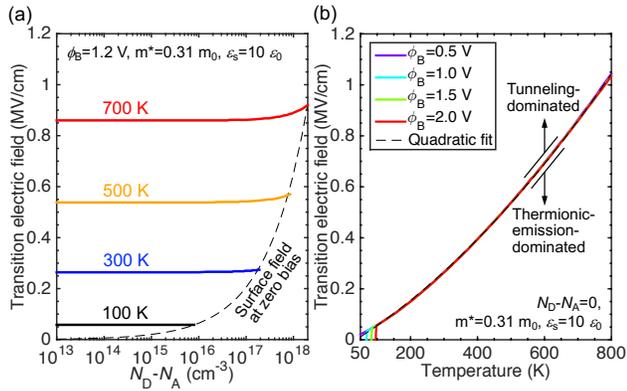

Fig. 2. Calculated transition electric field ($E_T$) in β-Ga$_2$O$_3$ SBDs as a function of (a) the net doping concentration ($N_D$-$N_A$) and (b) temperature.

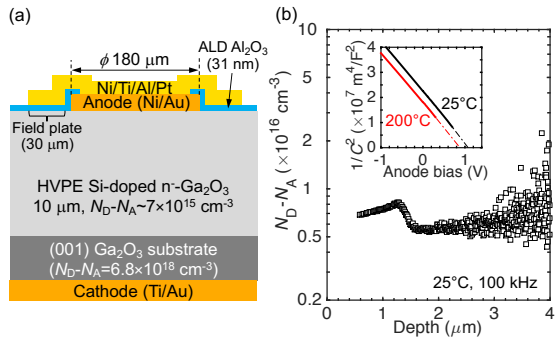

Fig. 3. (a) Schematic of the field-plated Ni-Ga$_2$O$_3$ SBDs fabricated on a HVPE Ga$_2$O$_3$ epitaxial wafer. (b) Extracted net doping concentration from $C$-$V$ measurements on SBDs without the field plate. Inset shows the $1/C^2$-$V$ plot which is used to extract the built-in potential zero bias ($V_{bi,0}$).

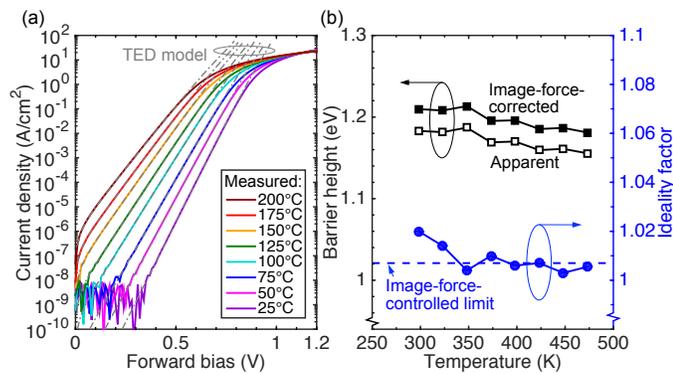

Fig. 4. (a) Temperature-dependent forward $I$-$V$ characteristics of the Ga$_2$O$_3$ SBDs, as well as the fitting using the thermionic emission-diffusion (TED) model. (b) Extract barrier heights (apparent and image-force corrected values) as well as ideality factors as a function of temperature.

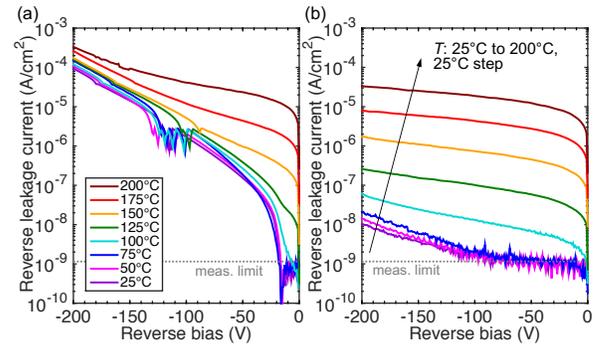

Fig. 5. Temperature-dependent reverse $I$-$V$ characteristics on the Schottky barrier diodes (a) without the field plate, and (b) with the field plate.

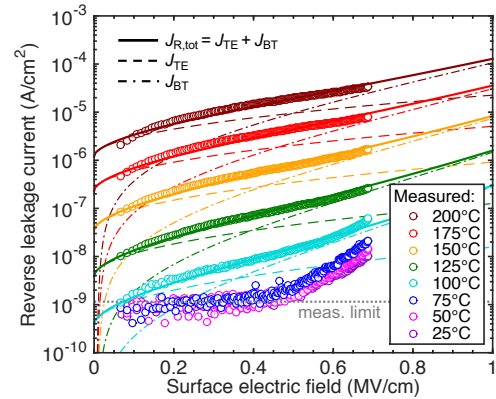

Fig. 6. Temperature-dependent reverse leakage current as a function of the surface electric field ($J$-$E$ characteristics) in the field-plated Ga$_2$O$_3$ SBDs. The data is fitted with the calculated total reverse leakage current ($J_{R,tot}$) *with the barrier height as the only fitting parameter*. The constituent components, $J_{TE}$ and $J_{BT}$, are also shown.

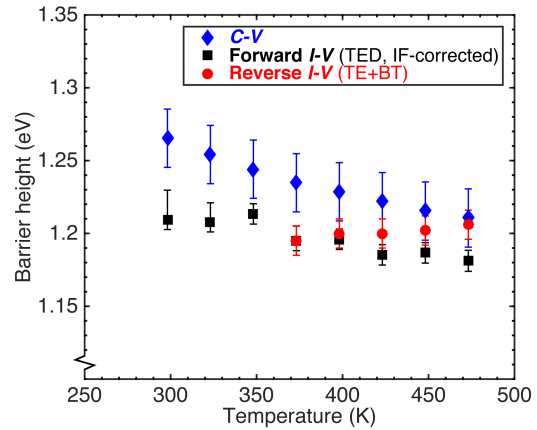

Fig. 7. Extracted barrier heights from $C$-$V$ measurements, forward $I$-$V$ measurements (using TED model) and reverse $I$-$V$ measurements (using our numerical leakage model considering both TE and BT).